\documentclass[preprint,12pt]{elsarticle}

\usepackage{amsmath,amssymb}
\usepackage[colorlinks=true, citecolor=blue, linkcolor=black, urlcolor=blue ]{hyperref}

\newcommand{\Eqn}[1]{Eq.~(\ref{#1})}

\newcommand\sss{\scriptscriptstyle}
\newcommand\as{\alpha_{\sss S}}
\newcommand\aW{\alpha_{\sss W}}
\newcommand\gs{g_{\sss S}}
\newcommand\gW{g_{\sss W}}
\newcommand\pt{p_{\sss T}}

\journal{Physics Letters B}

\begin{document}

\begin{frontmatter}

\title{Single-top $t$-channel production with off-shell and non-resonant effects}

\author[DESY]{A.S. Papanastasiou}
\author[CERN]{R. Frederix}
\author[CERN,EPFL]{S. Frixione}
\author[EPFL]{V. Hirschi}
\author[CP3-Louvain]{F. Maltoni}
\address[DESY]{DESY, Deutsches Elektronen-Synchrotron, Notkestra{\ss}e 85, \\ D-22607 Hamburg, Germany}
\address[CERN]{PH Department, TH Unit, CERN, CH-1211 Geneva 23, Switzerland}
\address[EPFL]{ITPP, EPFL, CH-1015 Lausanne, Switzerland}
\address[CP3-Louvain]{Centre for Cosmology, Particle Physics and Phenomenology (CP3), \\ 
Universit\'e catholique de Louvain, B-1348 Louvain-la-Neuve, Belgium}

\begin{abstract}
This letter details and discusses the next-to-leading order QCD corrections to
$t$-channel electro-weak $W^+ b j$ production, where finite top-width effects
are consistently taken into account. The computation is done within the 
{\sc aMC@NLO} framework and includes both resonant and non-resonant 
contributions as well as interferences between the two. Results are presented
for the LHC and compared to those of the narrow-width approximation and
effective theory approaches.
\end{abstract}


\end{frontmatter}

\section{Introduction}
\label{sec:intro}
Since its observation at the Tevatron in 2009
\cite{Aaltonen:2009jj,Abazov:2009ii}, single-top production has played an
important role in top-quark phenomenology at hadron colliders despite 
having a smaller cross section than that of $t\bar{t}$. This channel is not
only an important background to other processes, such as top-pair and Higgs
production, but it is also interesting in its own right, for example for its
potential to allow a direct determination of the CKM matrix element $V_{tb}$ 
(see \cite{Lacker:2012ek} for a recent discussion on this point), 
as well as being sensitive to new physics effects in many beyond-the-Standard
Model scenarios.  Single-top production is also a process through which the
measurements of key top quark properties such as the mass, $m_t$, can be
performed independently from those in $t\bar{t}$ production. This is 
particularly useful because the majority of the systematic uncertainties 
are expected to be fairly different in these two processes.

It is of no surprise therefore that over the last three decades much effort
has been invested in providing accurate predictions for single-top cross
sections.  Stable
single-top production was first discussed at the leading order (LO) in
\cite{Willenbrock:1986cr}. Next-to-leading order (NLO) QCD corrections in the
five-flavour (5F) scheme were first computed in \cite{Harris:2002md} and in
the four-flavour (4F) scheme in \cite{Campbell:2009ss,Campbell:2009gj}, whilst
electro-weak (EW) corrections were investigated in
\cite{Beccaria:2008av}. Furthermore, soft and collinear gluon resummation has
been studied in
\cite{Kidonakis:2010tc,Kidonakis:2011wy,Zhu:2010mr,Wang:2010ue} and stable
single-top at NLO matched to parton showers has been achieved in
\cite{Frixione:2005vw,Alioli:2009je,Frederix:2012dh} for both the 5F 
and 4F schemes.  Incorporating the decay
of the top quark, $t\rightarrow W^+ b$, as part of the hard process matrix
elements was accomplished via the narrow-width approximation (NWA) in
\cite{Campbell:2004ch,Heim:2009ku,Schwienhorst:2010je}, in which NLO corrections to both production and decay
were included. 
Going beyond the NWA, a study of off-shell and non-resonant effects at LO was 
performed in \cite{vanderHeide:2000fx} and final-state non-factorizable 
corrections in the $s$-channel were examined in \cite{Pittau:1996rp}.
In \cite{Falgari:2011qa,Falgari:2010sf},
effective theory (ET) techniques were employed to relax the assumption of an
on-shell top quark in the amplitudes, allowing for a systematic study of finite
top-width ($\Gamma_t$) effects in the resonant regions of phase space.

The full computation of $\alpha_s$-corrections to EW $W^+ b j$ production
(i.e. the process that includes single-top production) has so far been missing
in the literature. Given the opportunity for precision top physics provided by
the LHC, it is of phenomenological interest to understand the effects of
off-shell top quarks versus top quarks in the NWA. In this letter we present
such a computation for the $t$-channel process in the 5F scheme,
and make a direct comparison to results in the NWA as well as to those 
obtained by using the ET method.  The $s$-channel and $Wt$-production 
processes are not considered here.

The expectation is that for inclusive observables off-shell effects are 
small \cite{Fadin:1993dz,Fadin:1993kt,Melnikov:1993np} (i.e., parametrically
of $\mathcal{O}(\Gamma_t/m_t)$), whilst they should be noticeable 
in the case of less inclusive observables, such as the invariant or
transverse masses of the reconstructed top. Indeed, differences between 
NWA and ET or off-shell calculations have already been highlighted for 5F 
scheme single-top \cite{Falgari:2011qa,Falgari:2010sf} and top-pair
\cite{Bevilacqua:2010qb,Denner:2012yc,Denner:2010jp,Falgari:2013gwa}
production. We shall confirm these findings here.

This paper is organized as follows. In Sect.~\ref{sec:unstable}
we briefly discuss the NWA, ET, and off-shell approaches in view of
their application to single-top production; results for total rates
and for a few selected distributions are presented in Sect.~\ref{sec:res}; 
we conclude in Sect.~\ref{sec:conclusions}.

\section{Unstable single-top production}
\label{sec:unstable}
In the NWA limit $\Gamma_t \rightarrow 0$ the following replacement 
is made in the squared amplitude,
\begin{align}
\frac{1}{(p^2_t-m^2_t)^2 + \Gamma^2_t \, m^2_t} \;\longrightarrow\;
\frac{\pi}{m_t \Gamma_t} \delta(p^2_t-m^2_t)\,.
\label{eq:NWA}
\end{align}
The error introduced by making this approximation is expected to be of
order $\Gamma_t/m_t$.  The replacement above leads to
an exact factorization of the matrix elements into terms describing the
production and the decay of on-shell top quarks.  This factorization can be
combined with a strict expansion of $\Gamma_t$ in $\alpha_s$ to yield results
correct to $\mathcal{O}(\alpha_s)$ \cite{Melnikov:2009dn,Campbell:2012uf}.

Maintaining a finite width and subsequently relaxing the on-shell assumption
in standard fixed-order perturbation theory entails the computation of much
more than the resonant diagrams (i.e., those that feature an $s$-channel 
top-quark propagator) 
required in the NWA. Namely, at any given
perturbative order one should include all the resonant {\em and} non-resonant
diagrams (the latter feature a $W$-boson and a $b$ quark, which are not
connected to each other in a $tWb$ vertex) in order not to break gauge
invariance.

In the region where the invariant mass of the $Wb$ pair is close, but 
not necessarily equal to $m_t$, the cross-section is dominated by 
resonant diagrams. It is possible to make use of this fact to
construct an expansion of the matrix elements around the pole of the 
top-quark propagator \cite{Stuart:1991cc,Aeppli:1993rs}.  The pole expansion
has been generalized through the use of ET techniques in
\cite{Falgari:2011qa,Falgari:2010sf,Falgari:2013gwa}, where applications to
the processes of single-top and top-pair production were also made, thus 
capturing the dominant finite-width effects.

The disadvantage of both the NWA and ET approaches is that they are in
principle only valid in regions of phase space where the top quarks are either
on or near their mass-shell. A gauge-invariant way of introducing a
finite-width whilst ensuring that the final result is valid over the full
phase space is the complex-mass scheme (CMS) \cite{Denner:1999gp,Denner:2005fg}, 
whereby the
top-quark width is consistently included at the level of the Lagrangian via
the procedure of renormalization. The CMS is a generalization of the on-shell
renormalization scheme and has been recently employed by two groups to perform
NLO calculations of unstable top-pair production
\cite{Bevilacqua:2010qb,Denner:2012yc,Denner:2010jp}.  For the case of
unstable top quarks, it amounts to expressing the bare mass in terms of a
complex renormalized mass and a complex counter-term,
\begin{align}
m_{t,0} = \mu_t + \delta\mu_t,
\end{align}
where $\mu^2_t = m^2_t - i m_t \Gamma_t$. As is shown explicitly in
\cite{Denner:2005fg}, the quantity $\delta\mu_t$ can be fixed in terms of the
renormalized top-quark self-energy evaluated at the complex argument, $p^2_t =
\mu^2_t$, such that $\mu^2_t$ corresponds to the complex pole of the top quark
propagator. The precise value of the top width can be freely chosen as an
input in this scheme; but in order to ensure NLO accuracy, the width correct
to (at least) $\mathcal{O}(\alpha_s)$ should be used.

The CMS has recently been implemented \cite{Hirschi:CMS} in the framework of 
{\sc aMC@NLO}, and the results presented in this paper illustrate the first
hadron-collider application of this new feature. The automation of such an
approach to unstable particle production and decay is highly beneficial due to
the non-trivial book-keeping involved in these calculations.  The NLO
corrections have thus been obtained in an automated fashion, with the one-loop
and real contributions computed using {\sc MadLoop} \cite{Hirschi:2011pa} and 
{\sc MadFKS} \cite{Frederix:2009yq} respectively.

\subsection{Process definition}
\begin{figure}[t!]
\centering
\includegraphics[trim=4.8cm 20.6cm 1.2cm 3.5cm,clip,width=9.5cm]{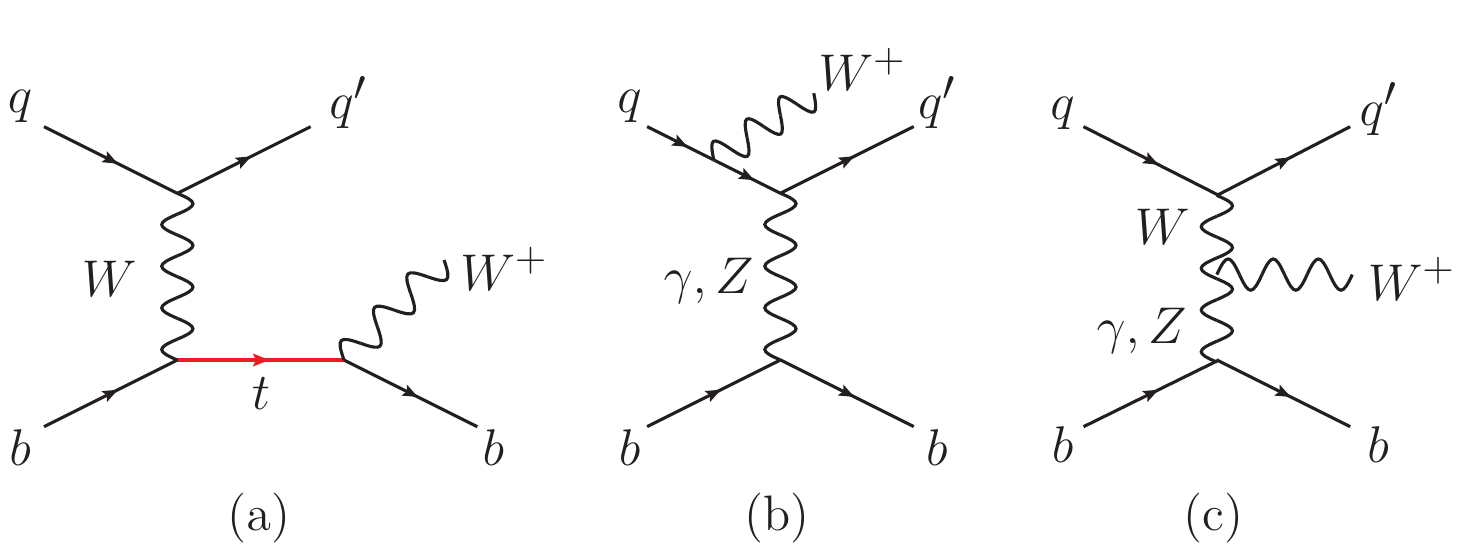}
\caption{Selection of LO $t$-channel diagrams for EW $W^+ b j$ production in the 5F scheme: resonant (a) and non-resonant (b) \& (c).}
\label{fig:lo-tch-diagrams}
\end{figure}

\begin{figure}[t!]
\centering
\includegraphics[trim=4.8cm 20.5cm 1.0cm 3.5cm,clip,width=9.5cm]{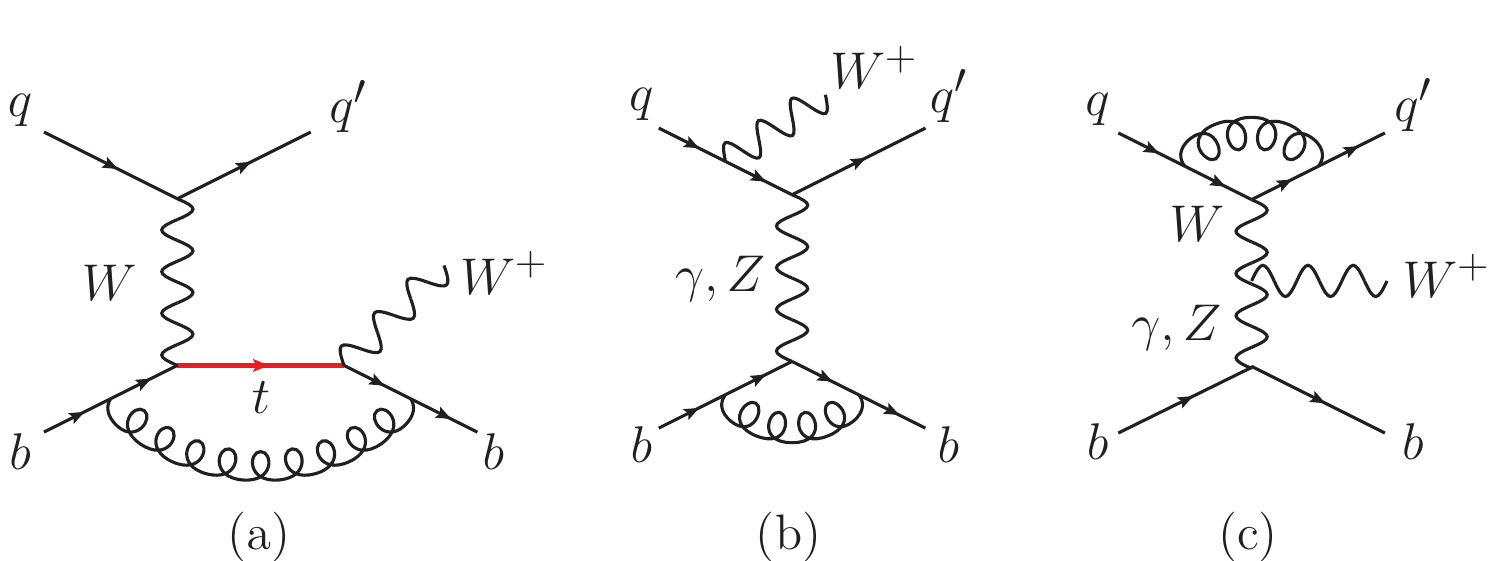}
\caption{Selection of NLO virtual $t$-channel diagrams for EW $W^+ b j$ production in the 5F scheme: resonant (a) and non-resonant (b) \& (c).}
\label{fig:nlo-tch-diagrams}
\end{figure}

\begin{figure}[t!]
\centering
\includegraphics[trim=4.8cm 21.8cm 0.9cm 4.2cm,clip,width=9.5cm]{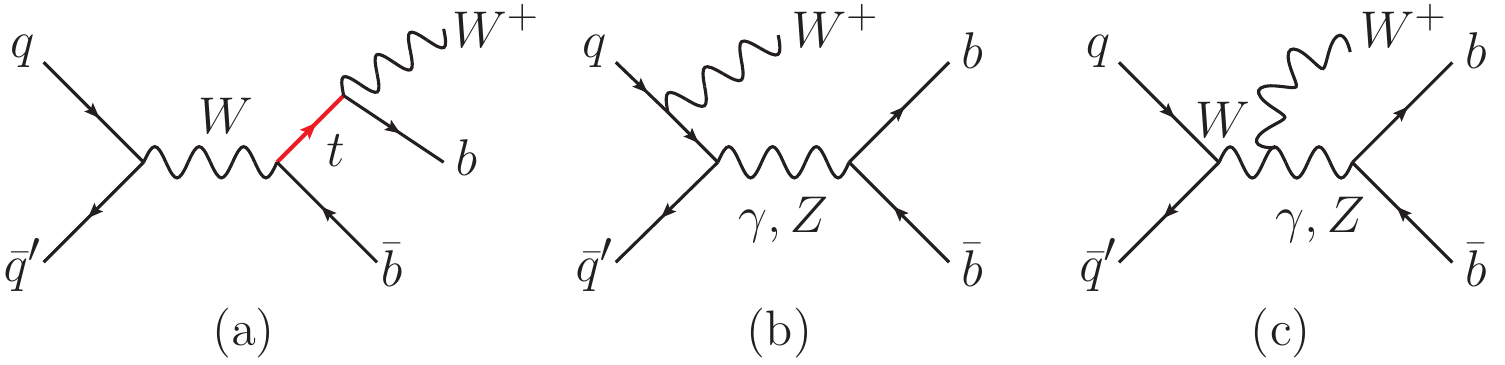}
\caption{Selection of LO $s$-channel diagrams for EW $W^+ b j$ production: resonant (a) and non-resonant (b) \& (c). These are not
included in our calculation.}
\label{fig:lo-sch-diagrams}
\end{figure}

Given that our aim is that of comparing the predictions of the
NWA and ET approaches to single-top production with those we
obtain by retaining all $\Gamma_t$ and interference effects,
the process we consider is:
\begin{align}
p\, p\, &\rightarrow \, W^+ \, J_b \, J_{\text{light}} + X.
\label{eq:stable-w-proc} 
\end{align} 
$J_b$ is defined to be a jet that contains at least a $b$ quark, while
$J_{\text{light}}$ is a jet that may (or may not) contain $b$
or $\bar{b}$ quarks.
Furthermore, at the Born level, we only keep contributions to the
cross-section of ${\cal O}(\aW^3)$. NLO QCD corrections to this set 
of diagrams are then computed, yielding contributions of ${\cal O}(\aW^3\as)$.
We stress that, while this is a straightforward procedure from
an algorithmic point of view when the top quark is on-shell, it becomes
highly non-trivial when this condition is relaxed. In particular, for 
the process at hand, the computation of NLO corrections in QCD is 
well defined {\em only} if the CKM matrix is diagonal in the third generation 
(i.e., $V_{tb}=1$). In fact, when $V_{tb}\ne 1$, Born-level amplitudes of
${\cal O}(\gW\gs^2)$ (which feature a gluon propagator) have a non-null
interference with those we consider here, thus resulting in a contribution of
${\cal O}(\aW^2\as)$ to the Born cross-section. The EW corrections to this
term are therefore of ${\cal O}(\aW^3\as)$, which is the same order as the NLO
QCD corrections to pure-EW Born matrix elements. One would therefore have no
choice but to compute both types of correction\footnote{Having said that, we
note that these interference terms are strongly CKM-suppressed, since one
needs two off-diagonal CKM matrix elements that connect a $b$ with a first-
or second-generation quark.}. In order to avoid such complications we will
not consider the case $V_{tb}\ne 1$ any further.

A consequence of having a third-generation diagonal CKM is that, up to NLO in
QCD and in the 5F scheme, $t$- and $s$-channel contributions to $Wbj$ 
production do not interfere (note that this is true in the case of on-shell 
top production regardless of the nature of the CKM matrix). We can thus
safely adopt the procedure of simply excluding $s$-channel Feynman diagrams,
keeping only those that feature the exchange of $t$-channel EW vector
bosons (samples of the latter diagrams are depicted in Figures
\ref{fig:lo-tch-diagrams} and \ref{fig:nlo-tch-diagrams}, and of the
former diagrams in Figure \ref{fig:lo-sch-diagrams}). In fact, at the LO
the identities of the partons in $t$-channel diagrams can never be
equal to those entering $s$-channel ones. This is not the case at the NLO,
where the two channels can however be simply distinguished by looking
at their colour structures. Discarding $s$-channel diagrams has the advantage 
of allowing us to directly compare with $t$-channel results available in
the literature. 

Even when setting $V_{tb}=1$, there is another important difference with
respect to the case of an on-shell top quark. Namely, while the latter does
not need any Born-level kinematic cuts to be well defined, this is not the
case for the full off-shell process, because of the presence of
potentially singular contributions from non-resonant diagrams -- see for
example Figure \ref{fig:lo-tch-diagrams}b, where the photon propagator
diverges when the outgoing $b$ quark is parallel to the beam line.  This
implies that it is necessary to impose a constraint on the final-state 
$b$-jet for our CMS computation to be well defined.
We can summarise the discussion so far as follows: when $V_{tb}=1$
the $t$-channel contribution to the process of \Eqn{eq:stable-w-proc} and its
NLO QCD corrections are finite and well defined, provided that $J_b$ has a
non-zero transverse momentum.

While the $s$-channel cross section is strongly suppressed at the LHC, one may
wonder whether it can be competitive with the non-resonant effects in the
$t$ channel. Although we shall not study this problem here, we
note that it is possible to include both $s$ and $t$ channels in our approach,
provided we slightly modify the definition of $J_b$ to being a jet that 
contains one $b$ quark, but not a $b\bar{b}$ pair (or in other words, $J_b$
cannot be $b$-flavour neutral). This definition is in fact sufficient to damp
singularities in diagrams that feature a $\gamma\to b\bar{b}$ splitting
(which, being of QED origin, are not subtracted in an NLO QCD computation).
As in the previous case, a minimal $\pt(J_b)$ must be imposed.

\section{Results}
\label{sec:res}
The minimal conditions necessary to ensure that the process of
\Eqn{eq:stable-w-proc} is well defined have been given above.  In our
phenomenological analysis, we supplement these with additional kinematic cuts,
which are summarised in Table \ref{table:cuts}. In this way, similarly to the
comparison made recently in \cite{AlcarazMaestre:2012vp} between NWA
\cite{Bernreuther:2004jv,Melnikov:2009dn,Campbell:2012uf} and off-shell
\cite{Bevilacqua:2010qb,Denner:2012yc,Denner:2010jp} approaches for $t\bar{t}$
production and decay, we ensure not to impose conditions on our final state
that could potentially enhance non-resonant contributions.
This allows for a fair comparison to be made with the NWA and ET results
available in the literature.  As far as the NWA is concerned, we use the
$t$-channel single-top production and decay process implemented in MCFM
\cite{Campbell:2004ch,Campbell:2012uf};
for the ET approach we have used the $t$-channel single-top code 
of \cite{Falgari:2010sf}. In addition to the decay of the top
quark, the NWA and ET results we compare to also include the decay $W^+
\rightarrow e^+ \nu_e$. To ease the comparison we have therefore included a
factor of $1/9$ in our off-shell results to account for the appropriate
$W$-boson branching fraction. However, no information on the decay products 
of the $W$-boson is used in the analysis we perform.

\begin{table}
\centering
\begin{tabular}{c c}
\hline \\[-5pt]
$p_T(J_b) > 25 \text{ GeV}$ & $p_T(J_{\text{light}}) > 25 \text{ GeV} $ \\[5pt]
$\lvert \eta(J_b) \rvert < 4.5$ & $\lvert \eta(J_{\text{light}}) \rvert < 4.5$ \\[5pt]
\multicolumn{2}{c}{$140 < M(W^+,J_b) < 200 $ GeV} \\[5pt]
\hline
\end{tabular}
\caption{Basic analysis setup allowing for a fair comparison of NWA, ET and off-shell approaches.}
\label{table:cuts}
\end{table}

\begin{table}
\centering
\begin{tabular}{c c}
\hline \\[-7pt]
$m_Z = 91.1876 \text{ GeV}$ & $m_W = 80.3980 \text{ GeV}$ \\[5pt]
$\Gamma_Z = 2.4952 \text{ GeV}$ & $\Gamma_W = 2.1054 \text{ GeV}$ \\[5pt]
$G_F=1.6639 \times 10^{-5} \text{ GeV}^{-2} $ & $\alpha^{-1}_{e}=132.3384$ \\[5pt]
\hline
\end{tabular}
\caption{Input parameters.}
\label{table:parameters}
\end{table}

The pole mass of the top quark is set equal to $m_t = 173.2$ GeV and 
we use the \texttt{MSTW2008NLO} PDF set \cite{Martin:2009iq} (at both
LO and NLO),
which also provides the value of $\alpha_s(m_Z)$. Further input parameters are
shown in Table \ref{table:parameters}.  The top quark widths we use at LO and
NLO are $\Gamma^{\text{LO}}_t=1.5017$ GeV and
$\Gamma^{\text{NLO}}_t(\mu=m_t/2)=1.3569$ GeV respectively.  
Jets are defined by means of the $k_t$-algorithm 
\cite{Catani:1993hr,Ellis:1993tq}, as
implemented in {\sc FastJet} \cite{Cacciari:2011ma}, with $R=0.5$.  
The central value for the factorization and renormalization scales
is $\mu=m_t/2$. These two scales are also simultaneously varied in 
the range $\mu \in [m_t/4, m_t]$, in order to estimate the size of
missing higher order corrections; we point out that while performing
these variations, the scale is also consistently changed in the top quark 
width.  The upper panels of the plots we present display the LO
(green band) and NLO (blue band) off-shell results, along with the NWA
(solid-red with crosses) and ET (solid-magenta with boxes) results. 
The lower insets show the relative differences between the various 
approaches (NWA/ET in  solid-red/dashed-magenta), by plotting the 
quantities 
$d\sigma^{\text{NWA/ET}}_{\text{NLO}}/ 
d\sigma^{\text{off-shell}}_{\text{NLO}} -1$.

\begin{table}[h!]
\centering
\begin{tabular}{c || c | c }
&  LO & NLO \\[2pt]
\hline \hline
& & \\[-7pt]
CMS [pb]  & $4.184(1)_{-12.3\%}^{+8.5\%}$ & $4.115(5)_{+4.6\%}^{+0.5\%}$ \\[5pt]
\hline
& & \\[-7pt]
NWA [pb] & $4.223(1)_{-12.2\%}^{+8.8\%}$ & $4.138(1)_{+2.6\%}^{+0.9\%}$ \\[5pt]
\%diff & +0.9 & +0.6 \\[2pt]
\hline
& & \\[-7pt]
ET [pb] & $4.154(1)_{-12.2\%}^{+8.8\%}$ & $4.074(1)_{+4.0\%}^{+0.3\%}$ \\[5pt]
\%diff & -0.7 & -1.0 \\[2pt]
\hline
\end{tabular}
\caption{LHC (8 TeV) cross sections for the process defined via the analysis
  of Table \ref{table:cuts}, at LO and NLO for the off-shell (CMS), NWA and ET
  computations.  Numbers in brackets are Monte Carlo integration uncertainties
  whilst the percentages indicate scale uncertainties. `\%diff' is the \%
  difference to the CMS results. }
\label{table:cross-sections}
\end{table}

\begin{figure}[t!]
\centering
\includegraphics[trim=1.2cm 0.2cm 3.0cm 0.3cm,clip,width=11.2cm]{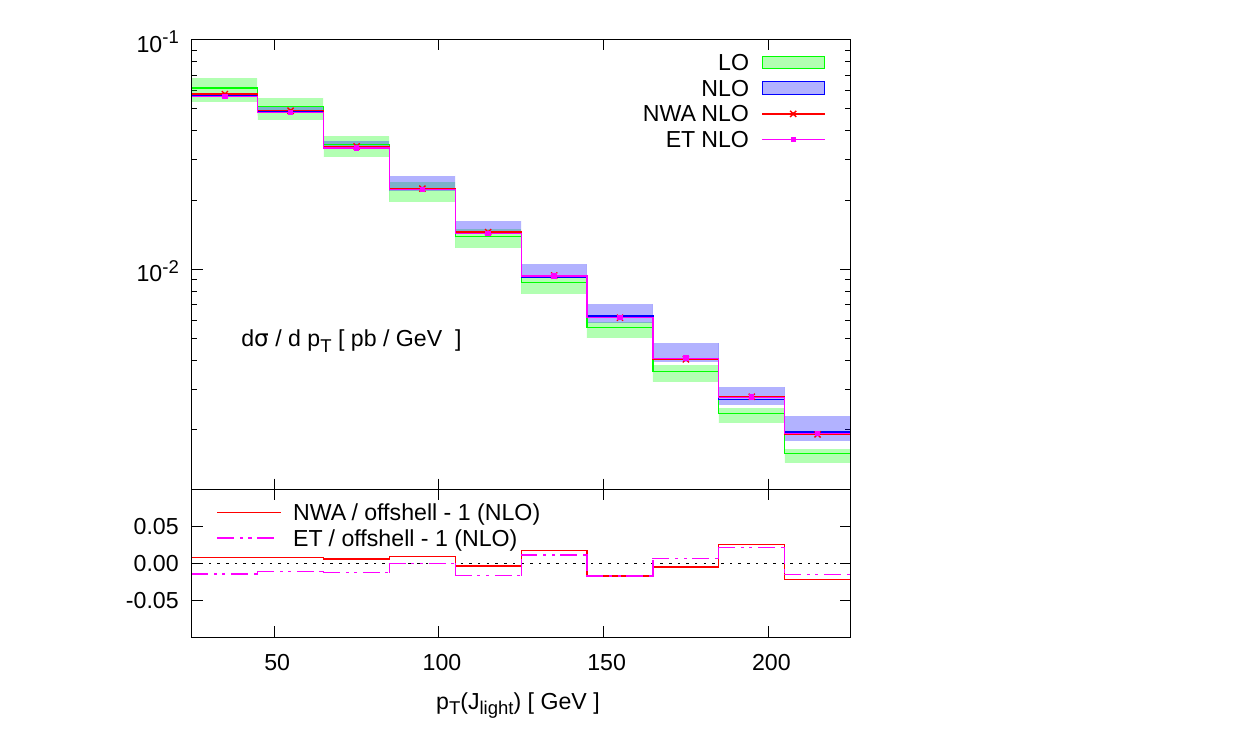}
\caption{Transverse momentum of light jet, $p_T(J_{\text{light}})$. }
\label{fig:pt-2ndjet}
\end{figure}

The cross sections at the LO and NLO resulting from our analysis in the
off-shell, NWA and ET approaches are catalogued in Table
\ref{table:cross-sections}.  The key point to highlight here is the small
difference, $\mathcal{O}(1\text{-}2\%)$, between the three approaches,
consistent with our expectation that it be parametrically
suppressed in the NWA by terms of $\mathcal{O}(\Gamma_t/m_t)$
for inclusive observables. Indeed, similar
small-sized differences are observed for differential observables either
inclusive in, or insensitive to, the invariant mass of the ($W^+,J_b$)-system.
As an illustrative example we present in Figure \ref{fig:pt-2ndjet}
the transverse momentum distribution of the light jet, 
$p_T(J_{\text{light}})$. The lower panel reveals that the NWA 
and ET NLO results differ by 1-2\% in all bins from the 
off-shell NLO results. In the upper panel it can
be seen that both the NWA and ET results are actually contained 
within the scale variation band of the NLO off-shell result, indicating 
that for this observable the size of off-shell effects is smaller than 
the scale uncertainty. 

\begin{figure}[h]
\centering
\includegraphics[trim=1.2cm 0.2cm 3.0cm 0.3cm,clip,width=11.2cm]{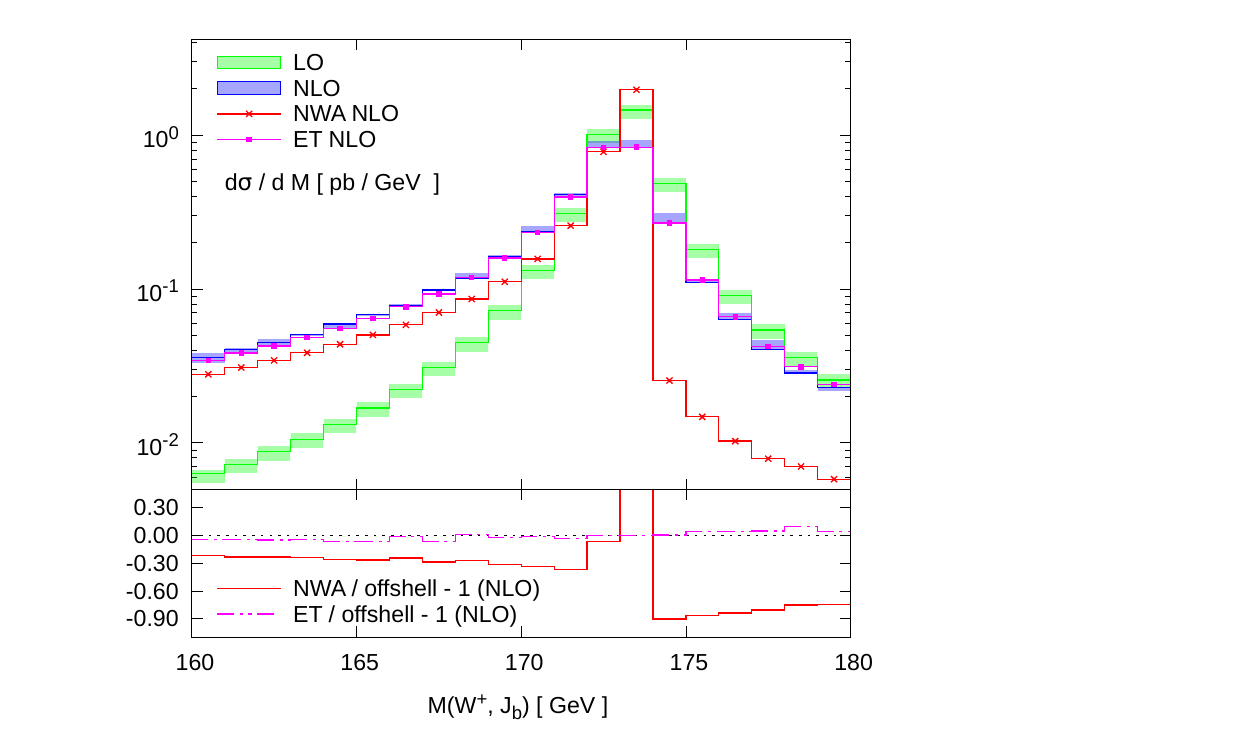}
\caption{Invariant mass distribution for the reconstructed top quark,
  $M(W^+,J_b)$.}
\label{fig:inv_mass}
\end{figure}

The picture changes for observables which are less inclusive in the invariant
mass of the reconstructed top quark (i.e., the $(W^+,J_b)$-system), with the
prime example being of course the invariant mass itself, displayed in Figure
\ref{fig:inv_mass}. The first feature one observes is that the NLO corrections
are large, in particular below the peak position. The origin of these is to a
large extent the real corrections to the top decay, confirmed by the
fact that the NWA result mimics the shape of the off-shell curve for
$M(W^+,J_b) < m_t$.  However, it is clear that the shapes of the distributions
predicted by the NWA and off-shell calculations are very different for
$M(W^+,J_b) > m_t$ where the NWA curve is much steeper than the one in which
finite-width effects have been included. This is due to the fact that in the
NWA the $W^+J_b$ invariant mass can only receive contributions 
to $M(W^+,J_b) > m_t$
from real NLO corrections, where the additional radiation originates in the
{\em production} process, and is then clustered into the $b$-jet. On the 
contrary, in the off-shell calculation, the $M(W^+,J_b)$ distribution 
receives contributions in this region already at the LO, both from
resonant diagrams (due to the off-shellness of the top quark), as well as from
non-resonant ones.  As indicated in the lower panel, near the peak of the
distribution and for larger mass values the difference between the 
full off-shell result and that in the NWA exceeds 60\% and is 
much bigger than scale uncertainty. Another important feature to pick out
from the lower panel is the change in sign of this difference at the
peak. This is the reason why, despite the sizeable differences between the
off-shell and NWA distributions seen in Figure \ref{fig:inv_mass}, the
off-shellness only results in small corrections
for inclusive
observables \cite{Falgari:2011qa,Falgari:2010sf}. In contrast to the case of
the NWA, the agreement between the off-shell and ET results is striking, over
the full range considered here.

\begin{figure}[t]
\centering
\includegraphics[trim=1.2cm 0.2cm 3.0cm 0.3cm,clip,width=11.2cm]{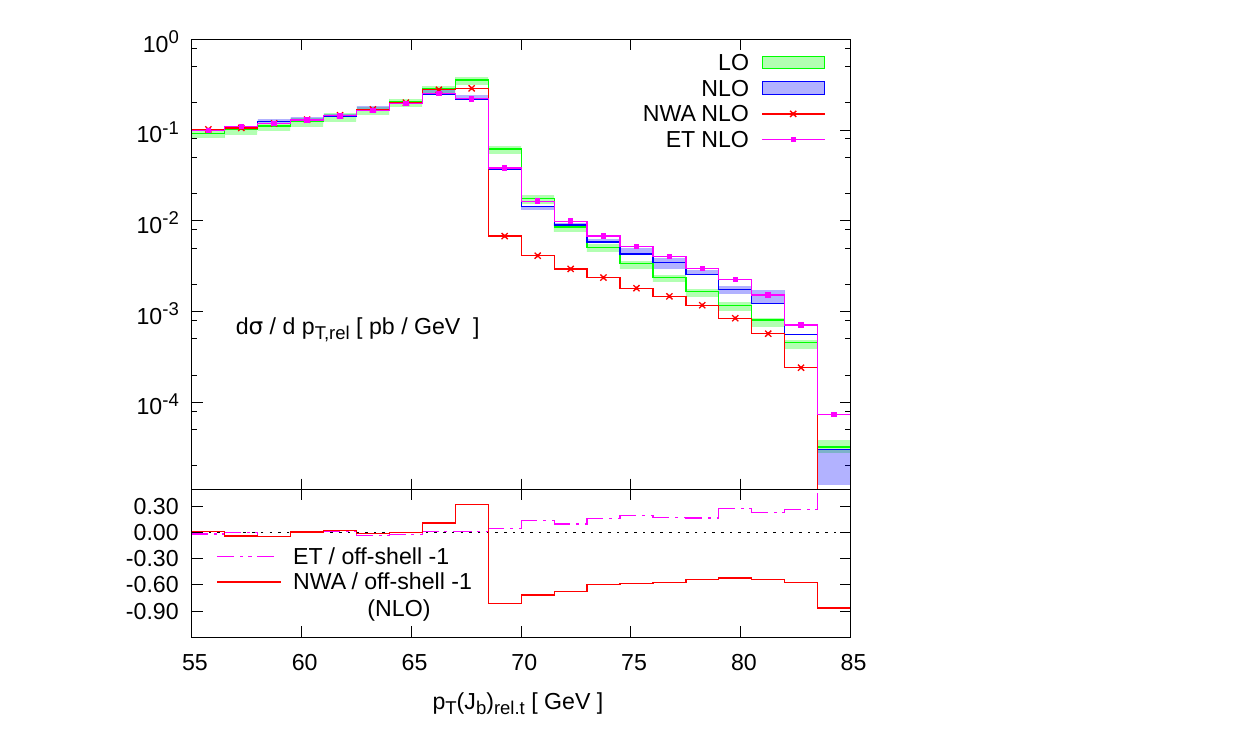}
\caption{Transverse momentum of $b$-jet relative to flight of top quark, in
  reconstructed top quark rest frame, $p_T(J_b)_{\text{rel.t}}$.}
\label{fig:rel_pt}
\end{figure}

We now examine Figure \ref{fig:rel_pt}, where we plot the distribution of
transverse $b$-jet momentum relative to the direction of flight of the
reconstructed top quark in the reconstructed top-quark rest frame,
$p_T(J_b)_{\text{rel.t}}$.  At the LO and when the top quark is assumed to be
on-shell this observable has a kinematic cut-off at $p_T(J_b)_{\text{rel.t}} =
(m^2_t-m^2_W)/(2m_t)$.  Once again, the difference between the NWA and
off-shell approaches changes sign at the sharp edge of the distribution, and
beyond it finite-width effects can be very large. Both the NWA and ET results 
approximate the off-shell one well in the region below the edge. Close
to the peak and beyond it, the NWA is in vast disagreement with the off-shell
calculation, as was the case for the invariant-mass distribution.
The ET prediction gives a rather good description at the peak; beyond
the edge the ET and off-shell are still relatively close to each other
(i.e., almost within scale variation), but there is
an obvious trend in the difference between the two.
This is an indication that, towards the upper end of the range considered, a
region of phase space is probed in which non-resonant, or
subleading-$\Gamma_t$, contributions become important. For the same reason,
similar differences between the ET and CMS results will show up in the tails
of the invariant mass distribution should the range considered there be
enlarged.
Figures \ref{fig:inv_mass} and \ref{fig:rel_pt} underline that the use of such
observables in certain contexts (e.g., template fitting) 
requires care when using NWA predictions.  Finally, we stress
that these findings are consistent with and complement those
discussed in recent studies of finite-width effects in top quark production
\cite{Falgari:2011qa,Falgari:2010sf,Bevilacqua:2010qb,Denner:2012yc,
Denner:2010jp,Falgari:2013gwa}.

\section{Conclusions}
\label{sec:conclusions}
In this letter we have performed the computation of NLO QCD corrections to EW
$t$-channel $W^+ b j$ production. The calculation, carried out within the
{\sc aMC@NLO} framework, was done making use of the complex-mass scheme, and 
retains the full off-shell and interference effects at NLO. In addition we
have compared our results with those obtained with the NWA and ET approaches.
We conclude that, at least in the case of the top quark, 
it is incorrect to claim that the NWA is an excellent
approximation universally. While the NWA gives a good description of many
observables, it fails dramatically for others, in particular those sensitive
to the invariant mass of the ($W^+, J_b$)-system. On the other hand, we find
that the predictions of the ET approach are much closer to those of the full
NLO QCD results.  These two facts combined imply that for certain
observables off-shell effects are much more relevant for a correct
description of the final-state kinematics, than NLO corrections to the
top-quark decay alone (which include hard radiation off the $b$ quark). We
feel that this is a general conclusion and should be applicable
regardless of the hard process. This has
already been confirmed by similar results reported in the literature for
$t\bar{t}$ production.  Clearly, along with the dominant off-shell effects,
there are process-dependent and beyond-leading-$\Gamma_t$ effects, which one
can start to see in the small differences between the full NLO QCD and ET
results. These differences, which are numerically subleading for the present
analysis, are expected to become larger if the non-resonant effects are 
allowed to become important, for example by requiring the invariant 
mass of ($W^+, J_b$)-system to be significantly larger than $m_t$.

An important future study will be that of assessing the impact that parton
showers and hadronization have on the distributions presented here.  Closely
connected to this, possible systematics on the top-mass extraction introduced
by ignoring finite-width effects at the matrix element level can be
quantified for single-top for a realistic experimental setup.

\section*{Acknowledgements}
\label{sec:acknowledgements}
We thank the authors of \cite{Falgari:2010sf} for providing us with their code
for the ET results.  AP thanks Paolo Torrielli for his help with {\sc aMC@NLO}
and gratefully acknowledges the support and hospitality of the CERN Theory
division where much of this work was completed.  This research has been
supported by the ERC grant 291377 ``LHCtheory: Theoretical predictions and
analyses of LHC physics: advancing the precision frontier."
SF is on leave of absence from INFN, Sez. di Genova.

\section*{References}

\bibliographystyle{utphys_cs}
\bibliography{wbj_prod_refs}

\end{document}